\documentclass{aastex}  
\usepackage{natbib}
\usepackage{emulateapj5}
\usepackage{apjfonts}
\shorttitle{Apparent magnitudes in an inhomogeneous universe}
\begin{document}  
\title{Apparent magnitudes in an inhomogeneous universe: the global
viewpoint}
\author{Hamish G. Rose}
\affil{University of Canterbury, Christchurch, New Zealand}
\email{h.rose@phys.canterbury.ac.nz}

\begin{abstract}
Apparent magnitudes are important for high precision cosmology.  It is
generally accepted that weak gravitational lensing does not affect the
relationship between apparent magnitude and redshift.  By considering
metric perturbations it is shown that objects observed in an inhomogeneous
universe have, on average, higher apparent magnitudes than those observed
at the same redshift in a homogeneous universe.
\end{abstract}
\keywords{gravitational lensing --- galaxies: distances and redshifts --- cosmology: observations --- distance scale}
\section{Introduction}

Implicit in many modern cosmological experiments, such as the Supernova
Cosmology Project \citep{perlmutter98} and the High-z Supernova Search
\citep{schmidt98}, is the assumption that the relationship between the
apparent magnitude of a distant object and the redshift of that object
is unaffected by gravitational lensing.

\citet{weinberg76} presents an argument that gravitational lensing does
not, on average, affect the apparent magnitude -- radial Robertson
Walker coordinate relationship.  His argument was based on photon
number conservation and has been widely accepted, although recently
it has come under closer scrutiny.  \citet{ellis98} have reanalysed
Weinberg's argument and concluded that it is not valid if caustics
are present, which only occurs with strong fields.  \citet{claudel00}
considers Newtonian perturbations in the weak field limit and finds that,
to first order in $\kappa = 8\pi G/c^2$, there is no deviation from the
Friedmann Robertson Walker (FRW) result.

Weinberg's original argument is presented in Section \ref{sec_coord}.
In Section \ref{sec_z} I demonstrate using metric perturbations that
the presence of inhomogeneities in the universe do, on average, affect
the apparent magnitude -- redshift relationship.

\section{Photon number count versus coordinate distance}
\label{sec_coord}

Consider an exactly FRW universe containing a source at (comoving
coordinate) $r=0$ that emits $N$ photons (due to, say, a cataclysmic
event).  Drawing a sphere around the source at $r=r_{\mathrm{obs}}$ (with surface
area $4\pi a_{\mathrm{obs}}^2r_{\mathrm{obs}}^2$ where $a(t)$ is the cosmological scale
factor), an astronomer at $r=r_{\mathrm{obs}}$ with a telescope of area $A$
will observe $n$ photons satisfying
\begin{equation}
\frac{4\pi a^2_{\mathrm{obs}}r_{\mathrm{obs}}^2}{A} = \frac{N}{n}.
\label{eq_photon_number}
\end{equation}

Now consider the above situation with the matter inside the sphere
distributed unevenly, thus lensing the photons. The number of photons
observed by the astronomer may be different when compared to the previous
situation. However, the total number of photons passing through the sphere
is unchanged.  If the area of the sphere at $r=r_{\mathrm{obs}}$ has not changed
then any increase or decrease in photons seen by the astronomer must
be compensated by a decrease or increase respectively in the number
of photons observed by other astronomers.  Furthermore, if there are
a large number of astronomers at different points on the sphere at
$r=r_{\mathrm{obs}}$ then the average number of photons they observe must be
distributed about $n$ with a standard deviation that approaches $0$
as their combined telescopes cover the sphere.

Weinberg's argument is based on the assumption that the area of the
sphere centred on $r=0$ and with radius $r=r_{\mathrm{obs}}$ is not affected by
the mass distribution.  The area of the sphere depends not only on $r$
but also on $a(t)$.  If inhomogeneities affect the time photons take to
travel from $r=0$ to $r=r_{\mathrm{obs}}$ then inhomogeneities must also affect
the area of the sphere.  If this is the case then photon conservation does
not imply that the observed apparent magnitude relationship is identical
to the FRW apparent magnitude relationship.  In Section \ref{sec_z}
this is shown to be the case.

\section{Apparent magnitude versus redshift}
\label{sec_z}

Consider a perturbed FRW dust universe with line element 
\begin{equation}
\mathrm{d}s^2 = c^2 \mathrm{d}t^2 - a(t)^2 (1-h(r,t))^2 ( \frac{ \mathrm{d}r^2}{1-kr^2} +r^2 \mathrm{d}\theta^2 + r^2\sin^2(\theta) \mathrm{d} \phi^2) 
\end{equation}
and an energy momentum tensor
\begin{equation}
T=\left[\begin{array}{cccc}
	  \rho_b (1+\delta(r,t)) & 0 & 0 & 0 \\
	  0 & 0 & 0 & 0 \\
	  0 & 0 & 0 & 0 \\  
	  0 & 0 & 0 & 0
\end{array}\right] ,	
\end{equation}
where $\rho_b$ is the matter density in the FRW universe and $\delta(r,t)$
describes the departure from homogeneity.  The coordinates are comoving and
peculiar motions are neglected so $T$ contains no terms dependent on
the velocity of matter.

The metric has determinant
\begin{equation}
\sqrt{-g} = a^3(1-h)^3.
\end{equation}

Spherical symmetry has been retained so that any astronomer at
$r=r_{\mathrm{obs}}$ makes the same observations of a source at
$r=0$ (as in Section \ref{sec_coord}) as any other astronomer at
$r=r_{\mathrm{obs}}$.  Spherical symmetry is an unnatural condition
to impose upon inhomogeneities.  However, no use is made here of the
matter distribution except in that no observer is in a special location.
In particular no dynamics are considered so spherical symmetry is a
reasonable condition to impose.

Conditions are imposed upon the perturbed universe to ensure that it does
not depart too far from FRW.  Specifically, the total mass content is the
same as in a FRW universe, which means that the function $a(t)$ is the
same in both cases; and the inhomogeneities are small in both amplitude
($|\delta| \ll 1$) and length.  In every region small enough that the
scale factor, $a(t)$, changes little in the time taken for the photon to
travel through it, $\delta$ averages to $0$.  Thus we impose the condition
\begin{equation}
\int_\lambda \frac{\delta(r(\lambda), t(\lambda))}{a(t(\lambda))} \mathrm{d}\lambda =0 
\label{eq_average}
\end{equation}
where $\lambda$ is any parameterization along the geodesic.  Finally,
$h(r,t)=0$ at both the source and the observer's locations since the only
effect under consideration here is due to matter inhomogeneities between
the source and the observer.

Following \citet[page 276]{peebles93} (but note the different definition
of $h$ which allows the calculation to be carried out to all orders),
the stress energy conservation law leads to
\begin{equation}
\frac{1}{\sqrt{-g}}\partial_\mu (\sqrt{-g} T^\mu_0) =\frac{1}{2}g_{\mu\nu,0}T^{\mu\nu} = 0
\end{equation}
or
\begin{equation}
\frac{\dot{\rho}_b}{\rho_b} + \frac{\dot{\delta}}{1+\delta} = -\frac{3\dot{a}}{a}+\frac{3\dot{h}}{1-h}.
\end{equation}

Since $\dot{\rho}_b / \rho_b = -3\dot{a} / a$ in the unperturbed universe,
\begin{equation}
\frac{\dot{\delta}}{1+\delta} = \frac{3\dot{h}}{1-h} 
\end{equation}
which has solution
\begin{equation}
1-h=(1+\delta)^{-\frac{1}{3}} .
\label{eq_def_h}
\end{equation}
The integration constant has been determined by requiring that
when $\delta=0$, $h=0$.

By integrating along a radial null geodesic from
emission at $(t,r)=(t_{\mathrm{em}},0)$ to observation at
$(t,r)=(t_{\mathrm{obs}},r)$, the radial coordinate where a photon arrives
at the observer in the perturbed universe can be determined and compared
to the radial coordinate of the observer in the FRW universe.

In the perturbed FRW universe 
\begin{equation}
\mathrm{d}s^2 = 0 \Rightarrow c\mathrm{d}t = a(t)(1-h(r,t)) \frac{\mathrm{d}r}{\sqrt{1-kr^2}} .
\label{eq_pert_de}
\end{equation}
Equation (\ref{eq_pert_de}) gives a differential equation which may be
solved for $r(t)$. The position of the photon is completely described
by the function $r(t)$.  $h(t)$ is now defined along the geodesic as
$h(t)\equiv h(r(t),t)$.  Similarly, $\delta(t)\equiv \delta(r(t),t)$.
Rearranging equation (\ref{eq_pert_de}) and integrating, one obtains
\begin{equation}
\int_{t_{\mathrm{em}}}^{t_{\mathrm{obs}}} \frac{c \mathrm{d}t}{a(t)(1-h(t))} = \int_{0}^{r_{\mathrm{obs}}} \frac{\mathrm{d}r}{\sqrt{1-kr^2}},
\label{eq_pert_geodesic_1}
\end{equation}

Similarly, in the FRW universe 
\begin{equation}
\int_{t_{\mathrm{em}}}^{t_{\mathrm{obs}}} \frac{c \mathrm{d}t}{a(t)} = \int_{0}^{r_{\mathrm{FRW}}} \frac{\mathrm{d}r}{\sqrt{1-kr^2}}.
\label{eq_FRW_geodesic}
\end{equation}

Substituting for $1-h$ from equation (\ref{eq_def_h}) and
making the second order approximation $(1+\delta)^{1/3} \approx
1+\delta/3-\delta^2/9$, equation (\ref{eq_pert_geodesic_1})
becomes
\begin{equation}
\int_{t_{\mathrm{em}}}^{t_{\mathrm{obs}}} \frac{c \mathrm{d}t}{a(t)} + \frac{1}{3}\int_{t_{\mathrm{em}}}^{t_{\mathrm{obs}}} \frac{c \delta(t)\mathrm{d}t}{a(t)} -\frac{1}{9}\int_{t_{\mathrm{em}}}^{t_{\mathrm{obs}}} \frac{c \delta(t)^2\mathrm{d}t}{a(t)} = \int_{0}^{r_{\mathrm{obs}}} \frac{\mathrm{d}r}{\sqrt{1-kr^2}}.
\label{eq_pert_geodesic_2}
\end{equation}

Equations (\ref{eq_FRW_geodesic}) and (\ref{eq_pert_geodesic_2}) are both
written using comoving coordinates and proper time and so may be compared
directly.  The first term in equation (\ref{eq_pert_geodesic_2}) may be
replaced using equation (\ref{eq_FRW_geodesic}) and the second term is $0$
due to equation (\ref{eq_average}) leaving
\begin{equation}
\frac{1}{9}\int_{t_{\mathrm{em}}}^{t_{\mathrm{obs}}} \frac{c \delta(t)^2\mathrm{d}t}{a(t)} = \int_{r_{\mathrm{obs}}}^{r_{\mathrm{FRW}}} \frac{\mathrm{d}r}{\sqrt{1-kr^2}}.
\label{eq_result}
\end{equation}

Since $\delta^2 \ge 0$ and $a > 0$ then $r_{\mathrm{FRW}} \ge r_{\mathrm{obs}}$ and $r_{\mathrm{obs}} =
r_{\mathrm{FRW}}$ only if $\delta=0$ at all points along the lightcone.  We may
use this to compare the observed number of photons in the perturbed
universe and the FRW universe.  From equation (\ref{eq_photon_number}),
\begin{equation}
\frac{n_{\mathrm{obs}}}{n_{\mathrm{FRW}}} = \frac{NA}{4\pi r_{\mathrm{obs}}^2 a(t_{\mathrm{obs}})^2} \frac{4\pi r_{\mathrm{FRW}}^2 a(t_{\mathrm{obs}})^2}{NA}
\end{equation}
\begin{equation}
 = \frac{r_{\mathrm{FRW}}^2}{r_{\mathrm{obs}}^2} \ge 1.
\end{equation}

Telescopes in a perturbed FRW universe receive on average more photons
from a source at a given redshift than telescopes with the same area in
a FRW universe and therefore have a higher apparent magnitude.  This is
the main result of this paper.

\section{Discussion}

The focusing theorem \citep[page 132]{SEF92} shows that a light beam
is magnified if it is affected by gravitational lensing but does not
go through a caustic.  In light of the focusing theorem, the result
in Section \ref{sec_z} is not surprising.  The conclusion of Section
\ref{sec_z} is significant because it shows that gravitational lensing
can cause magnification for all observers without violating conservation
of photon number.

The calculation in Section \ref{sec_z} is very general in that it
does not depend up the distribution of matter, only that there are matter
perturbations.  It does assume that the global structure of the universe is
FRW and that the departure from FRW is slight.  It seems plausible that
any greater departure from FRW will not remove the effect.  Furthermore,
the argument is based on the non-linearity of the relationship between
matter and the metric, so it is easy to see how it may applied in models
other than FRW.

\acknowledgements
I am grateful to L. Ryder, J. Adams, W. Joyce, S.Seunarine and S. Besier
for discussions and careful reading of drafts.

\end{document}